\documentclass[english,twocolumn,showpacs]{revtex4}

\usepackage{verbatim}
\usepackage{amsmath}

\usepackage{epsfig}
\usepackage{amsbsy}
\usepackage{amsfonts}
\usepackage{amssymb}
\usepackage{amsmath}
\usepackage{graphicx,color}
\usepackage{graphicx}
\usepackage{ulem}

\newcommand{\ket}[1]{| #1 \rangle}

\begin{document}
\title{Quantum computing with incoherent resources and quantum jumps}

\author{M.F. Santos$^{1}$, M. Terra Cunha$^{2}$, R. Chaves$^{3}$, A.R.R. Carvalho$^{4}$}

\affiliation{$^{1}$ Departamento de F\'isica, Instituto de Ci\^encias Exatas, Universidade Federal de Minas Gerais, Belo Horizonte, CP 702, 30123-970, Brazil}

\affiliation{$^{2}$Departamento de Matem\'atica, Universidade Federal de Minas Gerais, Belo Horizonte, CP 702, 30123-970, Brazil}

\affiliation{$^{3}$ ICFO-Institut de Ci\`encies Fot\`oniques, Mediterranean Technology Park, 08860 Castelldefels (Barcelona), Spain}

\affiliation{$^{4}$ Centre for Quantum Computation and Communication Technology, Department of Quantum Sciences, Research School of Physics and Engineering,
The Australian National University, Canberra, ACT 0200 Australia}

\begin{abstract}
Spontaneous emission and the inelastic scattering of photons are two natural processes usually associated with decoherence and the reduction in the capacity to process quantum information. Here we show that when suitably detected, these photons are sufficient to build all the fundamental blocks needed to perform quantum computation in the emitting qubits while protecting them from deleterious dissipative effects. We exemplify by showing how to teleport an unknown quantum state and how to efficiently prepare graph states for the implementation of measurement-based quantum computation.
\end{abstract}


\maketitle

In the traditional circuit model of quantum
computation~\cite{Nielsen:2000}, a quantum algorithm is
implemented by the sequential action of entangling
gates and local unitaries followed by a final measurement stage
that reveals the result of the computation. An alternative
approach is the measurement-based quantum computation
(MBQC)~\cite{Raussendorf:2001,Raussendorf:2003} where an initial
highly entangled multiqubit graph state is prepared and a
succession of adaptive measurements on the individual qubits
defines the implementation of a specific algorithm. In both
scenarios, quantum computation can be carried out with a basic
toolbox of three elements: measurements on the computational
basis, single qubit rotations, and specific two-qubit entangling
gates.

In both computational models, decoherence~\cite{Giulini:1996} poses a major obstacle to practical implementations as the qubits are encoded in systems that are unavoidably coupled to the environment. Emitted and scattered photons are two notable examples of natural processes associated with decoherence in quantum systems. However, when
detected, the same photons lead to the observation of quantum jumps~\cite{Carmichael:1993,Sauter:1986,Nagourney:1986,Gleyzes:2007}, which can be used to
read information out of the systems and to manipulate them to some extent.

Here we show that when the photons spontaneously emitted and inelastically scattered by the qubits into their surrounding environment are suitably monitored, the three elements of the quantum computing toolbox can be implemented. We also show
that the same scheme protects the system as a whole from dissipation while the computation is carried on.
We exemplify the scheme by showing how to teleport an
unknown quantum state, a protocol that already encompasses all the
required elements for any quantum computation. Furthermore, our method efficiently produces graph
states useful for the implementation of measurement-based quantum computation~\cite{Raussendorf:2001,Raussendorf:2003}.

\begin{figure}[h]
\includegraphics[width=8cm]{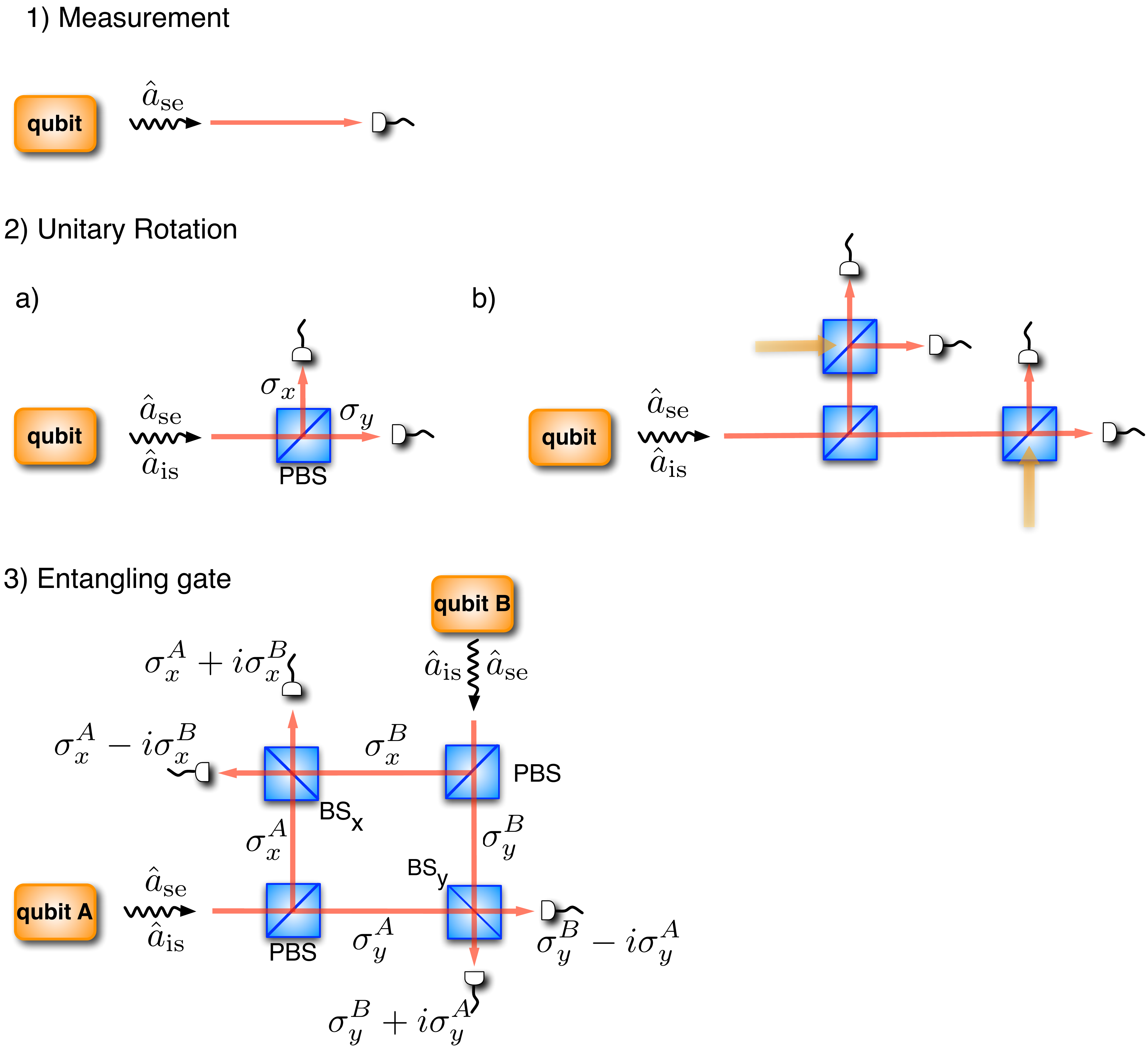}
\caption{Quantum-jump version of the building blocks for quantum computation. 1) Spontaneous  emission (``s.e.'') is sufficient for measuring in the computational basis. 2a) The PBS erases ``which process'' information and a photon-detection produces a jump proportional to a unitary flip in the emitting qubit. For $\theta=0$ the jumps correspond to $\sigma_x$ or $\sigma_y$~\cite{Carvalho:2011}; at other angles the jumps generate flips ($\pi$ rotations) around other directions in the equator of the Bloch sphere (linear combinations of $\sigma_x$ and $\sigma_y$). 2b) More general rotations can be obtained through homodyne detection schemes~\cite{Carvalho:2011}. 3) Entangling gates achieved through a second quantum erasing process: the standard balanced Beam Splitters (BS) erase ``which qubit'' information and a detection event in one of the available photocounters implements the corresponding maximally entangling operation indicated in the figure. After the first click, the standard BSs are removed and subsequent clicks will simply produce local flips as in (2a).}
\label{fig:detection}
\end{figure}

The first tool to be described is how to measure a qubit in the
computational basis what is promptly achieved by monitoring its
spontaneous emission (see Fig.~\ref{fig:detection}.1). If a non-degenerate qubit, prepared in a
superposition $\alpha|g\rangle + \beta|e\rangle$, is allowed to
decay at a rate $\gamma$, then, for times much larger than
$T=1/\gamma$, the excited (ground) component of the original state
of the qubit becomes correlated with the detection (no-detection)
of the spontaneously emitted photon. A detection identifies the
output $1$ (related to state $|e\rangle$) and defines a quantum
jump in the system that, mathematically, corresponds to applying the lowering operator $\sigma_-$ to the
state of the qubit. On the other hand, no-detection outputs $0$
and is associated to the so-called no-jump
trajectory~\cite{Carmichael:1993, Sauter:1986, Nagourney:1986,
Gleyzes:2007}. The monitoring of the spontaneous decay is then
equivalent to destructively measuring the state of the qubit in
the computational basis where, by destructive, we mean that the
measurement informs on the state of the qubit but always drags it
into its ground state.

\begin{figure}[h]
\includegraphics[width=8cm]{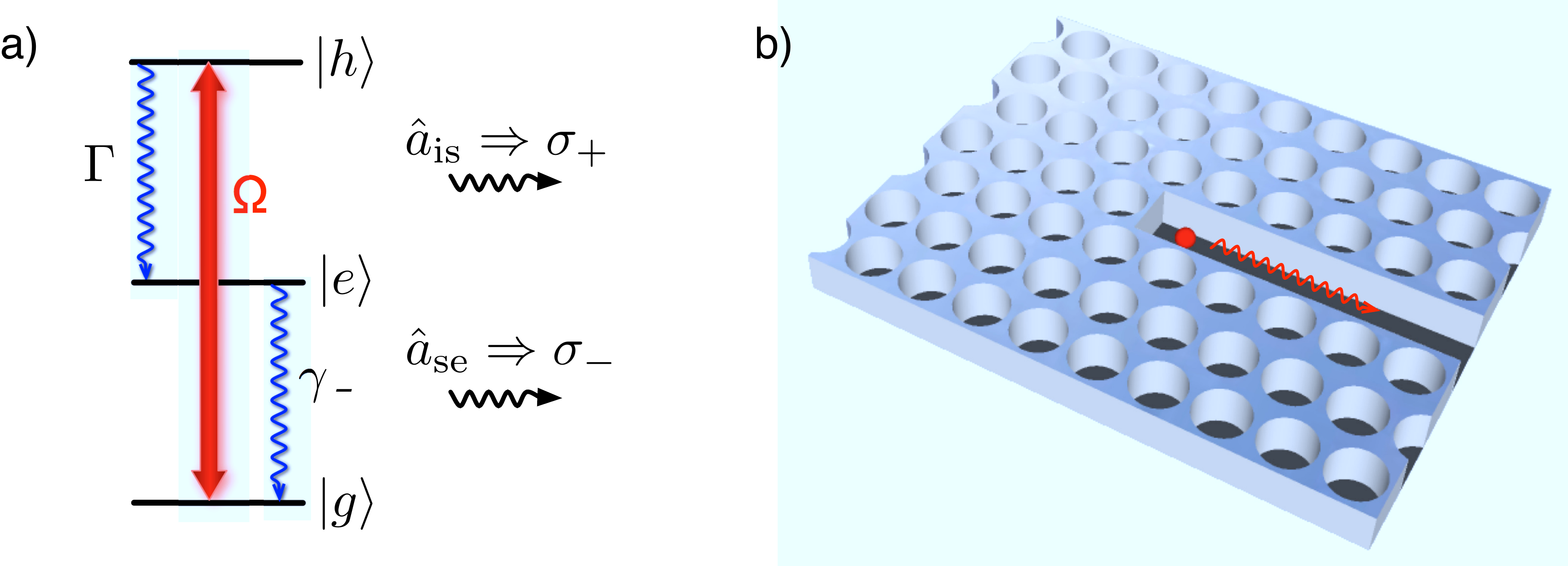}
\caption{Level scheme and emission collection. a) Each qubit is encoded in the two lower levels, $|g\rangle$ and $|e\rangle$ (corresponding to the logical $\ket{0}$ and $\ket{1}$) of a discrete emitter such as a quantum dot~\cite{Loss:1998}, a N-V center~\cite{Neumann:2008,Divincenzo:2010,Brouri:2000} or a superconducting qubit~\cite{Makhlin:2001,Vion:2002,Chiorescu:2003}. The third level, $|h\rangle$ is introduced to produce the necessary inelastic scattering that incoherently pumps the qubit back into its excited state $|e\rangle$ and complements
 spontaneous emission in order to generate the building blocks for quantum computation described in Fig.~\ref{fig:detection}. b) Efficient collection of the emission events can be obtained by placing the qubit inside or close to a 1-D system such as a photonic crystal~\cite{Barth:2009,Wolters:2010} (shown in the figure), a nanowire~\cite{Nadj-Perge:2010},  or some equivalent waveguide, in such a way that spontaneous emission and inelastic scattering happen in well defined directions. The output modes of this system (same spatial-temporal profile but orthogonal polarizations) are the source to the configurations of BSs and detectors shown in Fig.~\ref{fig:detection}.}
\label{fig:schemeatom}
\end{figure}

This ``destructive'' character means that the remaining tools for
quantum computation cannot be obtained solely by monitoring
spontaneously emitted (``s.e.'') photons. The curious solution
comes from the addition of an extra incoherent process to each
qubit: the inelastic scattering (``i.s.'') of photons provided by
an external laser field (see Fig.~\ref{fig:schemeatom}-a). When
this process optically pumps the emitting system back into its
excited state, the detection of photons in the ``i.s.'' channel
implements the complementary ``destructive'' jump $\sigma_+$ on
the qubit. However, if both ``s.e.'' and ``i.s.'' processes are
tuned to output photons that are indistinguishable in frequency
and linewidth but of orthogonal circular polarisations, then, by placing polarized beam splitters (PBS) before the photocounters,
the ``which process'' information is
erased~\cite{Scully:1982,Scully:1991}. In this case, shown in (Fig.\ref{fig:detection}.2-a), the detection of a
photon that comes out of the PBS will implement quantum jumps
that are linear combinations of $\sigma_-$ and $\sigma_+$ corresponding to unitary flips of the type $\cos \theta \sigma_x + \sin \theta \sigma_y$ on the qubit, with $\theta$ determined by the alignment of the PBS. More general
rotations are also possible if these new channels are monitored
through homodyne detection~\cite{Carvalho:2011}
(Fig.~\ref{fig:detection}.2-b) or combined with an external classical photon source.

The remaining building block, the two-qubit entangling gate, requires a second
quantum erasing process, one that destroys the ``which qubit''
information. By combining the output ports of the PBS of two
different qubits in a standard BS (see Fig.\ref{fig:detection}.3) information about the origin of the
photon (qubit A or B) is erased. As a consequence, clicks in
the detectors of these new channels will correspond to entangling
jumps given by linear combinations of local unitary flips, such as
$X_{AB}^{\pm}=(\sigma_{x_A}\pm i\sigma_{x_B})/\sqrt{2}$, where the sign is randomly determined by the
channel where the photon is detected. For example, if the initial state of the qubits is
$\ket{g_A}\otimes \ket{g_B}$, then a single detection event in
these combined detectors produces a maximally entangled Bell state
(Fig.\ref{fig:detection}.3).

After the first combined detection is obtained, the standard beamsplitters have to
be removed in order to avoid undoing the entangling operation.
In fact, this removal guarantees that any subsequent click in the same set of detectors will only
correspond to the application of a local flip in one of the qubits
thus preserving the existing entanglement shared by the
pair~\cite{Carvalho:2011}. This removal is now achievable in very
fast time scales, such as the method used in~\cite{Jacques:2007}
to implement Wheeler's delayed choice experiment. Also note that
the proposed entangling gate can be seen as an entanglement swap
operation between the propagating modes and the qubits.

\begin{figure}[h]
\includegraphics[width=8cm]{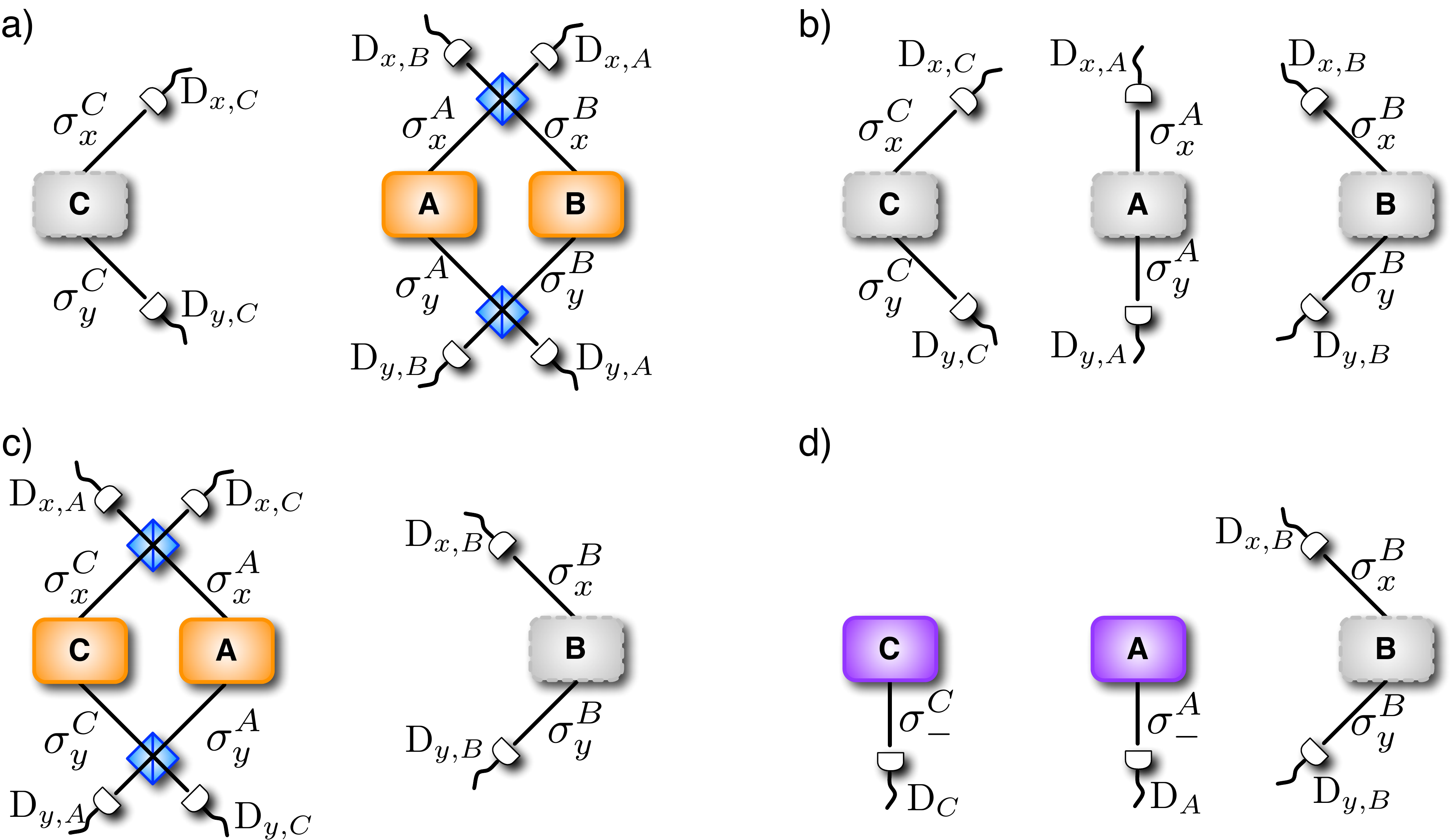}
\includegraphics[width=8cm]{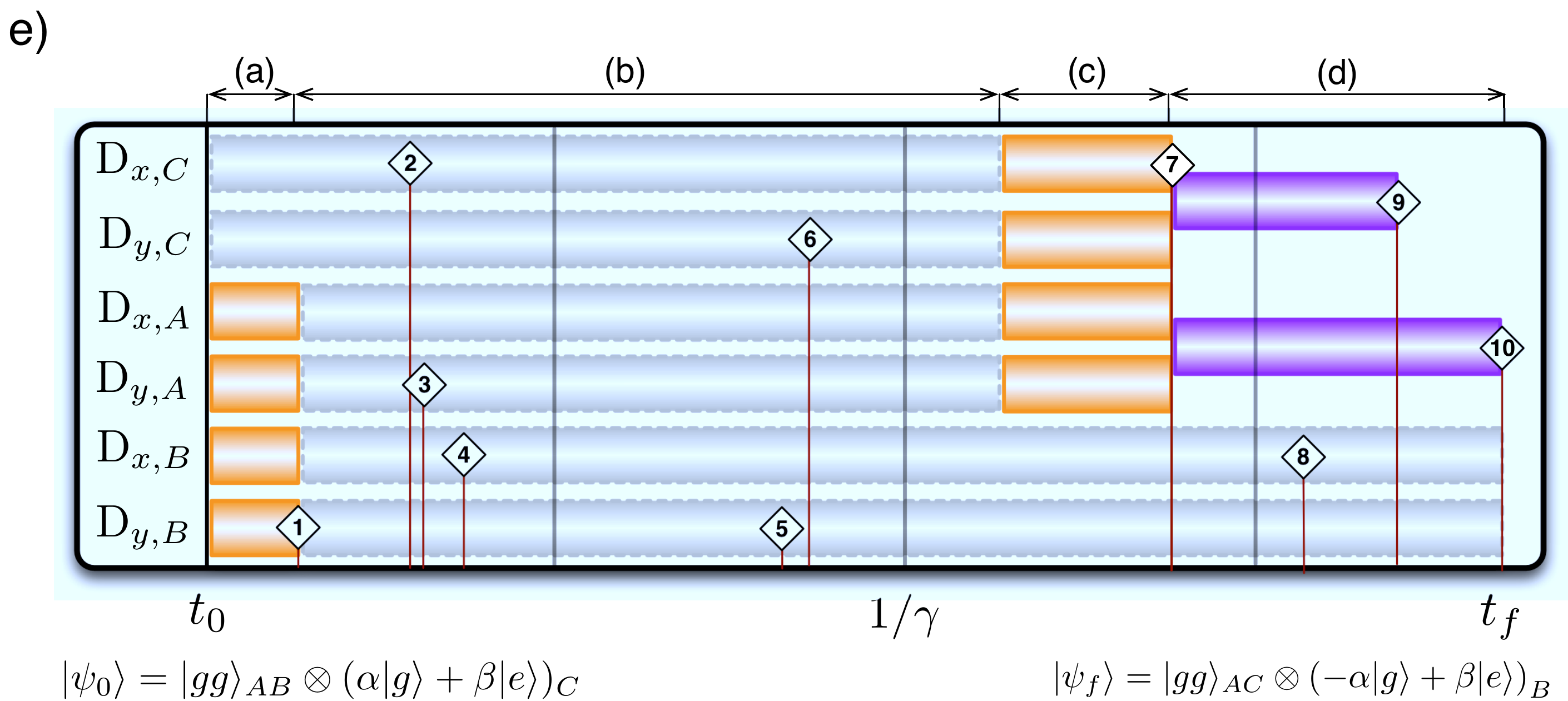}
\caption{Steps for the teleportation protocol. A  combined detection in $AB$ prepares Alice and
Bob's Bell pair (a). In (b) the beamsplitters are removed and only local flips can occur. An entangling operation between $A$ and $C$ is performed in (c) setting Alice's qubits  up for
computational basis measurements. With all the beamsplitters removed, and the ``i.s.'' mechanism turned off, qubits $A$ and $C$ are measured in (d). A simulation of these steps is shown in (e) with diamonds indicating the occurrence of a detection event.}
\label{fig:teleport}
\end{figure}

A major advantage of this gate over other ways to entangle qubits by monitoring spontaneously emitted photons~\cite{Jakobczyk:2002}
is its scalability. In fact, despite being based on detections, our gates are unitary, allowing the creation of arrays and networks
of entangled qubits. This property is essential, for example, in
the implementation of a teleportation protocol or the creation of
multiqubit graph states.

The teleportation steps are described in Fig.~\ref{fig:teleport}. In (a) Alice and Bob entangle their qubits using the
scheme shown in Fig.~\ref{fig:detection}.3. After a click in one of the detectors $D_{x(y),A(B)}$
the beam splitters must be removed, passing to configuration (b), where only entanglement-preserving
clicks (as in Fig.~\ref{fig:detection}.2-a) can be detected. The Bell measurement step starts in (c),
where Alice repeats the step in (a) but now entangling her qubit with Charlie's (the qubit in the state to be
teleported). After a click in one of the combined detectors, the beam splitters are removed and the driving fields
in Alice and Charlie's qubits turned off. In this situation, shown in (d), detection in $A$ and $C$ corresponds to
the scheme in Fig.~\ref{fig:detection}.1 and a measurement in the computational basis is performed. The diagram in
(e) represents the results of a simulation of the counts (shown by diamonds) in each detector for all the stages of the
teleportation protocol. The interval between steps (a) and (c) (the duration of step (b)) was arbitrarily set to $1/\gamma$. At the end of protocol, Alice and Charlie must communicate the results of their detections to Bob, which now, in possession
of the information about the measurements (in this case the 10 clicks detected), knows that the final state is $\ket{\psi_f}$
and he has to apply a simple Pauli operation (in this case $-\sigma_z$) to obtain Charlie's original state.

\begin{figure}
\includegraphics[width=8cm]{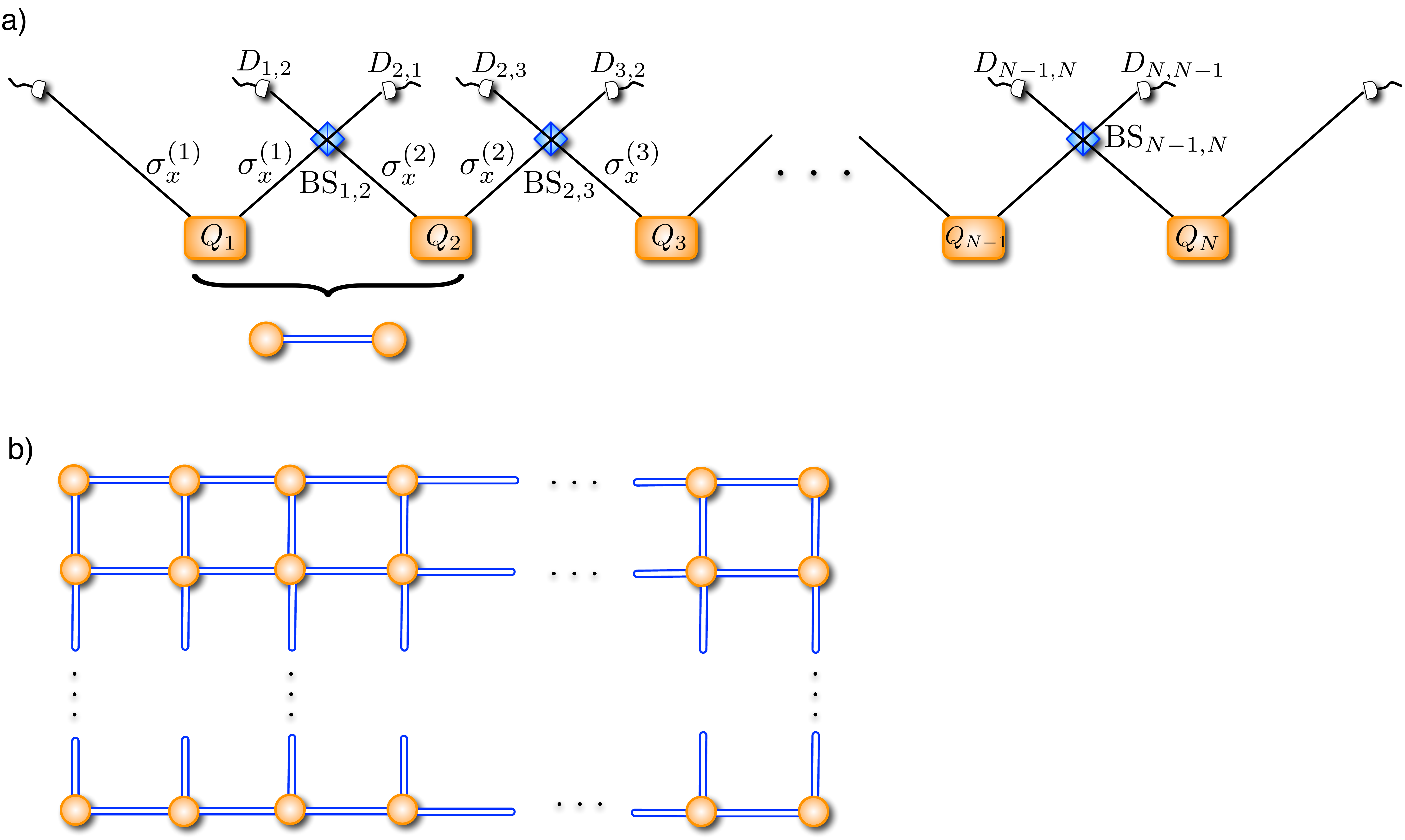}
\caption{Generation of graph states. a) The generation of a multiqubit state equivalent to a linear cluster state can be obtained by concatenating a variant of the entangling gate shown in Fig.~\ref{fig:detection}.3. In the case shown in the figure, only the $\sigma_x$ channels are combined at the BS while the $\sigma_y$ channels (not shown) should be detected independently (corresponding to the absence of ${\rm BS}_y$ in Fig.~\ref{fig:detection}.3). (b) With extra beam-splitters, one can connect a sequence of linear graphs to build a 2D cluster state, an universal resource for quantum computations.}
\label{fig:linear}
\end{figure}

The generation of a multiqubit graph state, as shown in
Fig.~\ref{fig:linear}, follows from the parallel application of
the entangling operation for every edge in the graph. All the
neighbour channels (in the graph topology) are initially combined.
However, since non-commuting operations are not desirable, only
one of the channels, for example $\sigma_x$, of each qubit should
be combined. The other channel is still monitored (corresponding to Fig.~\ref{fig:detection}.3 without $BS_y$) and its clicks
do not affect the overall result (they just correspond to locally redefining the logical ``$0$'' and ``$1$''). Each
time an entangling photon is detected, the corresponding standard
BS is removed and subsequent photons in the same detectors locally
flip the respective qubits.

For example, starting out with all $N$
qubits in the eigenstate $\vert 0 \rangle $ of $\sigma_z$\footnote{Note that since the entangling gates involve $\sigma_x$ operators, the qubits have
to be initialized in eigenstates of $\sigma_z$ or
$\sigma_y$ in order to form a proper graph state.}, the
generate state is
\begin{equation}
\vert \tilde{G}_N \rangle  = \prod_{(j,k)\in G} X_{jk}^{s_{j,k}} \ \vert 0 \rangle
^{\otimes N}
\label{eq:graph}
\end{equation}
where $s_{j,k}=\pm$ depends on which detector clicks in the
combined $BS_{j,k}$ channels and the $j$'s and $k$'s are chosen
accordingly to the connections present in the graph. Note that the $X_{jk}^{\pm}$ gates are formally equivalent (up to local unitaries) to Controlled-Z gates used to produce usual graph states as $\vert G_N \rangle = \prod_{(j,k)\in G} cZ_{j,k} \vert + \rangle ^{\otimes N}$. Indeed they can be rewriten as
$X_{jk}^{\pm} = e^{\pm i\pi/4}X^{\mp}_{j}X^{\pm}_{k}cX_{j,k}$, where $X^{\pm}_{j}=(\mathcal{I}\pm i\sigma_{x_j})$ is a local unitary and $cX_{j,k}$ is the $X$ basis version of the $cZ_{j,k}$ gate. Using that $H \sigma_{x_j} H = \sigma_{z_j}$ and the fact that all gates involving $\sigma _z$ commute, the state in Eq.~(\ref{eq:graph}) can then be rewritten as:
\begin{equation}
\vert \tilde{G}_{N}\rangle=  \prod _j U_j \vert G_N \rangle,
\end{equation}
with $ \prod _j U_j  =  H^{\otimes N} \prod_{(j,k)\in G} e^{s_{j,k}i\pi/4} Z^{-s_{j,k}}_{j} Z^{+ s_{j,k}}_{j}$,
$Z^{\pm s_{j,k}}_{j}=(\mathcal{I} \pm s_{j,k}i \sigma_{z_j})$ and $H$ being the usual Hadamard transformation. The state generated by our method is the usual graph-state, up to local unitaries conditioned on the clicks. It is worthy mentioning that all the measurements necessary to perform a universal one-way quantum computation are in the computational basis or comprised in the x-y plane of the Bloch sphere and can indeed be performed with the schemes presented in Fig.~\ref{fig:detection}.1 and Fig.~\ref{fig:detection}.2-a respectively. Given that, it can be formally shown that all the different clicks in the detectors (corresponding to different local unitaries) can be treated as the usual adaptations in the measurement basis of a given one-way algorithm. However, due to the products of Hadamards appearing in $U_j$ it is also necessary to introduce such transformations in our framework. This is easily achieved by the same quantum erasing principle that allows for the implementation of single qubit rotations in the x-y plane. However, in this case, an extra
classical photon source is required. For example, if the propagation channel of an attenuated laser is combined
in a standard beam splitter ($BS$) with the $\sigma_y$ output channel of a given emitting qubit $k$, then the detection of a photon
after the  $BS$ corresponds to applying the jump operator $(\mathcal{I} \pm i\sigma_{y_k})/\sqrt{2} = \exp \left( \pm i\frac{\pi}{4}\sigma_{y_k} \right)$ to qubit $k$, which also changes qubit $k$ from $Z$ basis to $X$ basis, and vice-versa.

The procedure to prepare graph states described above is efficient: If $\tau$ is the average time to entangle one pair (average time for one of two qubits to emit or
scatter one photon), then the parallel character of our proposal implies an overhead time of the order of $\tau \log N$ for the creation of an entire graph state comprised of $N$ edges. In order to prove this, note that the generation process is completed after the last combined detector has clicked, in an instance of the famous coupon collector problem~\cite{Durret:2010}. The mean value
for the time of occurrence of this last edge is of the order of
$\tau \log N$, where $\tau$ is the typical time for one
independent detection and $N$ is the number of edges on the graph.
As an example, for preparing a $2D$--cluster state of $100 \times
100$ qubits it will take about $12$ times the decay time of one
isolated qubit.

The results here presented are fundamentally different from other
proposals that use dissipation as resources for quantum
computation. There, either the interactions between qubits and
reservoirs have to be carefully and dynamically engineered
throughout the computation~\cite{Verstraete:2009, Barreiro:2011}, or
the protocol is probabilistic~\cite{Duan:2001, Olmschenk:2009}. Furthermore the deleterious effects of natural spontaneous
emission are always present. In contrast, here we consider local
reservoirs and incoherent pumping that always couple to each qubit
in the same way and the computation is obtained by suitably
observing the out-coming photons. Therefore, there is no need for
direct action neither on the qubits nor on the couplings and the
computation is naturally protected from spontaneous emission. Note also that in our case
 the process as a whole is stochastic but the computation is not probabilistic: each realization produces one of
many possible versions of the desired graph state, each version
differing from the others by local Pauli operations that can be
absorbed in the adaptation of measurement basis
required by a given computational algorithm.

We have investigated the computational power of quantum
trajectories and proved that one can implement any algorithm by
suitably observing the photons that are spontaneously emitted and inelastically
scattered by a multiqubit system. The results show
that a natural reservoir that destroys quantum
coherence when unobserved can be unravelled into quantum
trajectories with full quantum computational power.
From a fundamental point of view, one can interpret this
scheme as a careful and efficient selection of the set of quantum trajectories
that performs a desired computation out of the infinitude of possible
unravellings allowed for that given environment interaction. Our scheme suits both circuit and
measurement based quantum computation and we believe that recent
technological developments in $1$-D systems~\cite{Claudon:2010, Schoelkopf:2008, Hwang:2009}
will allow it to be tested in the near future.

The concepts here presented may also significantly contribute to
hybrid strategies to produce a quantum computer, for example, by
combining the here introduced entangling gates with more
traditional techniques to perform local operations. In that
respect, it is particularly encouraging the rather efficient way
to produce graph states by detecting the emitted and scattered
photons.

Other interesting directions of investigation are the integration
of the usual quantum error correcting codes in this computational
approach,  the development of new strategies to protect quantum
information in reservoir monitoring schemes, and the connection
with problems of percolation. Finally, the
intrinsic random nature of the clicks also indicates that the same
scheme could be used to generate random quantum circuits, $t$-designs
and quantum walks, all useful methods to study many-body quantum systems.

This work was conceived and majorly developed during the ``Quantum Information'' session of the ``Centro de Ciencias de Benasque Pedro Pascual'' in Benasque, Spain. MFS and MTC thank CNPq and Fapemig for financial support. R. C. was funded by the Q-Essence project. This work is part of the INCT-IQ from CNPq, Brazil. It was also supported by the Australian Research Council Centre of Excellence for Quantum Computation and Communication Technology (project number CE110001027).

\bibliography{$HOME/ARTICLES/allbib}

\begin{thebibliography}{30}
\expandafter\ifx\csname natexlab\endcsname\relax\def\natexlab#1{#1}\fi
\expandafter\ifx\csname bibnamefont\endcsname\relax
  \def\bibnamefont#1{#1}\fi
\expandafter\ifx\csname bibfnamefont\endcsname\relax
  \def\bibfnamefont#1{#1}\fi
\expandafter\ifx\csname citenamefont\endcsname\relax
  \def\citenamefont#1{#1}\fi
\expandafter\ifx\csname url\endcsname\relax
  \def\url#1{\texttt{#1}}\fi
\expandafter\ifx\csname urlprefix\endcsname\relax\def\urlprefix{URL }\fi
\providecommand{\bibinfo}[2]{#2}
\providecommand{\eprint}[2][]{\url{#2}}

\bibitem[{\citenamefont{Nielsen and Chuang}(2000)}]{Nielsen:2000}
\bibinfo{author}{\bibfnamefont{M.~A.} \bibnamefont{Nielsen}} \bibnamefont{and}
  \bibinfo{author}{\bibfnamefont{I.~L.} \bibnamefont{Chuang}},
  \textit{\bibinfo{title}{Quantum computation and quantum information}}
  (\bibinfo{publisher}{Cambridge University Press}, \bibinfo{year}{2000}).

\bibitem[{\citenamefont{Raussendorf and Briegel}(2001)}]{Raussendorf:2001}
\bibinfo{author}{\bibfnamefont{R.}~\bibnamefont{Raussendorf}} \bibnamefont{and}
  \bibinfo{author}{\bibfnamefont{H.~J.} \bibnamefont{Briegel}},
  \bibinfo{journal}{Phys. Rev. Lett.} \textbf{\bibinfo{volume}{86}},
  \bibinfo{pages}{5188} (\bibinfo{year}{2001}).

\bibitem[{\citenamefont{Raussendorf et~al.}(2003)\citenamefont{Raussendorf,
  Browne, and Briegel}}]{Raussendorf:2003}
\bibinfo{author}{\bibfnamefont{R.}~\bibnamefont{Raussendorf}},
  \bibinfo{author}{\bibfnamefont{D.~E.} \bibnamefont{Browne}},
  \bibnamefont{and} \bibinfo{author}{\bibfnamefont{H.~J.}
  \bibnamefont{Briegel}}, \bibinfo{journal}{Phys. Rev. A}
  \textbf{\bibinfo{volume}{68}}, \bibinfo{pages}{022312}
  (\bibinfo{year}{2003}).

\bibitem[{\citenamefont{Giulini and et~al.}(1996)}]{Giulini:1996}
\bibinfo{author}{\bibfnamefont{D.}~\bibnamefont{Giulini}} \bibnamefont{and}
  \bibinfo{author}{\bibnamefont{et~al.}}, \textit{\bibinfo{title}{Decoherence and
  the appearance of a classical world in quantum theory}}
  (\bibinfo{publisher}{Springer Press, Berlin}, \bibinfo{year}{1996}).

\bibitem[{\citenamefont{Carmichael}(1993)}]{Carmichael:1993}
\bibinfo{author}{\bibfnamefont{H.}~\bibnamefont{Carmichael}},
  \textit{\bibinfo{title}{An open systems approach to quantum optics}}, Lecture
  Notes in Physics m 18 (\bibinfo{publisher}{Springer-Verlag, Berlin},
  \bibinfo{year}{1993}).

\bibitem[{\citenamefont{Sauter et~al.}(1986)\citenamefont{Sauter, Neuhauser,
  Blatt, and Toschek}}]{Sauter:1986}
\bibinfo{author}{\bibfnamefont{T.}~\bibnamefont{Sauter}},
  \bibinfo{author}{\bibfnamefont{W.}~\bibnamefont{Neuhauser}},
  \bibinfo{author}{\bibfnamefont{R.}~\bibnamefont{Blatt}}, \bibnamefont{and}
  \bibinfo{author}{\bibfnamefont{P.~E.}~\bibnamefont{Toschek}},
  \bibinfo{journal}{Phys. Rev. Lett.} \textbf{\bibinfo{volume}{57}},
  \bibinfo{pages}{1696} (\bibinfo{year}{1986}).

\bibitem[{\citenamefont{Nagourney et~al.}(1986)\citenamefont{Nagourney,
  Sandberg, and Dehmelt}}]{Nagourney:1986}
\bibinfo{author}{\bibfnamefont{W.}~\bibnamefont{Nagourney}},
  \bibinfo{author}{\bibfnamefont{J.}~\bibnamefont{Sandberg}}, \bibnamefont{and}
  \bibinfo{author}{\bibfnamefont{H.}~\bibnamefont{Dehmelt}},
  \bibinfo{journal}{Phys. Rev. Lett.} \textbf{\bibinfo{volume}{56}},
  \bibinfo{pages}{2797} (\bibinfo{year}{1986}).

\bibitem[{\citenamefont{Gleyzes et~al.}(2007)\citenamefont{Gleyzes, Kuhr,
  Guerlin, Bernu, Deleglise, Busk~Hoff, Brune, Raimond, and
  Haroche}}]{Gleyzes:2007}
\bibinfo{author}{\bibfnamefont{S.}~\bibnamefont{Gleyzes}},
  \bibinfo{author}{\bibfnamefont{S.}~\bibnamefont{Kuhr}},
  \bibinfo{author}{\bibfnamefont{C.}~\bibnamefont{Guerlin}},
  \bibinfo{author}{\bibfnamefont{J.}~\bibnamefont{Bernu}},
  \bibinfo{author}{\bibfnamefont{S.}~\bibnamefont{Deleglise}},
  \bibinfo{author}{\bibfnamefont{U.}~\bibnamefont{Busk~Hoff}},
  \bibinfo{author}{\bibfnamefont{M.}~\bibnamefont{Brune}},
  \bibinfo{author}{\bibfnamefont{J.-M.} \bibnamefont{Raimond}},
  \bibnamefont{and} \bibinfo{author}{\bibfnamefont{S.}~\bibnamefont{Haroche}},
  \bibinfo{journal}{Nature} \textbf{\bibinfo{volume}{446}},
  \bibinfo{pages}{297} (\bibinfo{year}{2007}).

\bibitem[{\citenamefont{Carvalho and Santos}(2011)}]{Carvalho:2011}
\bibinfo{author}{\bibfnamefont{A.~R.~R.} \bibnamefont{Carvalho}}
  \bibnamefont{and} \bibinfo{author}{\bibfnamefont{M.~F.}
  \bibnamefont{Santos}}, \bibinfo{journal}{New Journal of Physics}
  \textbf{\bibinfo{volume}{13}}, \bibinfo{pages}{013010}
  (\bibinfo{year}{2011}).

\bibitem[{\citenamefont{Scully and Dr\"uhl}(1982)}]{Scully:1982}
\bibinfo{author}{\bibfnamefont{M.~O.} \bibnamefont{Scully}} \bibnamefont{and}
  \bibinfo{author}{\bibfnamefont{K.}~\bibnamefont{Dr\"uhl}},
  \bibinfo{journal}{Phys. Rev. A} \textbf{\bibinfo{volume}{25}},
  \bibinfo{pages}{2208} (\bibinfo{year}{1982}).

\bibitem[{\citenamefont{M.~O.~Scully}(1991)}]{Scully:1991}
\bibinfo{author}{\bibfnamefont{H.~W.} \bibnamefont{M.~O.~Scully},
  \bibfnamefont{Berthold-Georg~Englert}}, \bibinfo{journal}{Nature}
  \textbf{\bibinfo{volume}{351}}, \bibinfo{pages}{111} (\bibinfo{year}{1991}).

\bibitem[{\citenamefont{Loss and DiVincenzo}(1998)}]{Loss:1998}
\bibinfo{author}{\bibfnamefont{D.}~\bibnamefont{Loss}} \bibnamefont{and}
  \bibinfo{author}{\bibfnamefont{D.~P.} \bibnamefont{DiVincenzo}},
  \bibinfo{journal}{Phys. Rev. A} \textbf{\bibinfo{volume}{57}},
  \bibinfo{pages}{120} (\bibinfo{year}{1998}).

\bibitem[{\citenamefont{Neumann et~al.}(2008)\citenamefont{Neumann, Mizuochi,
  Rempp, Hemmer, Watanabe, Yamasaki, Jacques, Gaebel, Jelezko, and
  Wrachtrup}}]{Neumann:2008}
\bibinfo{author}{\bibfnamefont{P.}~\bibnamefont{Neumann}},
  \bibinfo{author}{\bibfnamefont{N.}~\bibnamefont{Mizuochi}},
  \bibinfo{author}{\bibfnamefont{F.}~\bibnamefont{Rempp}},
  \bibinfo{author}{\bibfnamefont{P.}~\bibnamefont{Hemmer}},
  \bibinfo{author}{\bibfnamefont{H.}~\bibnamefont{Watanabe}},
  \bibinfo{author}{\bibfnamefont{S.}~\bibnamefont{Yamasaki}},
  \bibinfo{author}{\bibfnamefont{V.}~\bibnamefont{Jacques}},
  \bibinfo{author}{\bibfnamefont{T.}~\bibnamefont{Gaebel}},
  \bibinfo{author}{\bibfnamefont{F.}~\bibnamefont{Jelezko}}, \bibnamefont{and}
  \bibinfo{author}{\bibfnamefont{J.}~\bibnamefont{Wrachtrup}},
  \bibinfo{journal}{Science} \textbf{\bibinfo{volume}{320}},
  \bibinfo{pages}{1326} (\bibinfo{year}{2008}).

\bibitem[{\citenamefont{DiVincenzo}(2010)}]{Divincenzo:2010}
\bibinfo{author}{\bibfnamefont{D.}~\bibnamefont{DiVincenzo}},
  \bibinfo{journal}{Nat Mater} \textbf{\bibinfo{volume}{9}},
  \bibinfo{pages}{468} (\bibinfo{year}{2010}).

\bibitem[{\citenamefont{Brouri et~al.}(2000)\citenamefont{Brouri, Beveratos,
  Poizat, and Grangier}}]{Brouri:2000}
\bibinfo{author}{\bibfnamefont{R.}~\bibnamefont{Brouri}},
  \bibinfo{author}{\bibfnamefont{A.}~\bibnamefont{Beveratos}},
  \bibinfo{author}{\bibfnamefont{J.-P.} \bibnamefont{Poizat}},
  \bibnamefont{and} \bibinfo{author}{\bibfnamefont{P.}~\bibnamefont{Grangier}},
  \bibinfo{journal}{Opt. Lett.} \textbf{\bibinfo{volume}{25}},
  \bibinfo{pages}{1294} (\bibinfo{year}{2000}).

\bibitem[{\citenamefont{Makhlin et~al.}(2001)\citenamefont{Makhlin, Sch\"on,
  and Shnirman}}]{Makhlin:2001}
\bibinfo{author}{\bibfnamefont{Y.}~\bibnamefont{Makhlin}},
  \bibinfo{author}{\bibfnamefont{G.}~\bibnamefont{Sch\"on}}, \bibnamefont{and}
  \bibinfo{author}{\bibfnamefont{A.}~\bibnamefont{Shnirman}},
  \bibinfo{journal}{Rev. Mod. Phys.} \textbf{\bibinfo{volume}{73}},
  \bibinfo{pages}{357} (\bibinfo{year}{2001}).

\bibitem[{\citenamefont{Vion et~al.}(2002)\citenamefont{Vion, Aassime, Cottet,
  Joyez, Pothier, Urbina, Esteve, and Devoret}}]{Vion:2002}
\bibinfo{author}{\bibfnamefont{D.}~\bibnamefont{Vion}},
  \bibinfo{author}{\bibfnamefont{A.}~\bibnamefont{Aassime}},
  \bibinfo{author}{\bibfnamefont{A.}~\bibnamefont{Cottet}},
  \bibinfo{author}{\bibfnamefont{P.}~\bibnamefont{Joyez}},
  \bibinfo{author}{\bibfnamefont{H.}~\bibnamefont{Pothier}},
  \bibinfo{author}{\bibfnamefont{C.}~\bibnamefont{Urbina}},
  \bibinfo{author}{\bibfnamefont{D.}~\bibnamefont{Esteve}}, \bibnamefont{and}
  \bibinfo{author}{\bibfnamefont{M.~H.} \bibnamefont{Devoret}},
  \bibinfo{journal}{Science} \textbf{\bibinfo{volume}{296}},
  \bibinfo{pages}{886} (\bibinfo{year}{2002}).

\bibitem[{\citenamefont{Chiorescu et~al.}(2003)\citenamefont{Chiorescu,
  Nakamura, Harmans, and Mooij}}]{Chiorescu:2003}
\bibinfo{author}{\bibfnamefont{I.}~\bibnamefont{Chiorescu}},
  \bibinfo{author}{\bibfnamefont{Y.}~\bibnamefont{Nakamura}},
  \bibinfo{author}{\bibfnamefont{C.~J. P.~M.} \bibnamefont{Harmans}},
  \bibnamefont{and} \bibinfo{author}{\bibfnamefont{J.~E.} \bibnamefont{Mooij}},
  \bibinfo{journal}{Science} \textbf{\bibinfo{volume}{299}},
  \bibinfo{pages}{1869} (\bibinfo{year}{2003}).

\bibitem[{\citenamefont{Barth et~al.}(2009)\citenamefont{Barth, N\"{u}sse,
  L\"{o}chel, and Benson}}]{Barth:2009}
\bibinfo{author}{\bibfnamefont{M.}~\bibnamefont{Barth}},
  \bibinfo{author}{\bibfnamefont{N.}~\bibnamefont{N\"{u}sse}},
  \bibinfo{author}{\bibfnamefont{B.}~\bibnamefont{L\"{o}chel}},
  \bibnamefont{and} \bibinfo{author}{\bibfnamefont{O.}~\bibnamefont{Benson}},
  \bibinfo{journal}{Opt. Lett.} \textbf{\bibinfo{volume}{34}},
  \bibinfo{pages}{1108} (\bibinfo{year}{2009}).

\bibitem[{\citenamefont{Wolters et~al.}(2010)\citenamefont{Wolters, Schell,
  Kewes, Nusse, Schoengen, Doscher, Hannappel, Lochel, Barth, and
  Benson}}]{Wolters:2010}
\bibinfo{author}{\bibfnamefont{J.}~\bibnamefont{Wolters}},
  \bibinfo{author}{\bibfnamefont{A.~W.} \bibnamefont{Schell}},
  \bibinfo{author}{\bibfnamefont{G.}~\bibnamefont{Kewes}},
  \bibinfo{author}{\bibfnamefont{N.}~\bibnamefont{Nusse}},
  \bibinfo{author}{\bibfnamefont{M.}~\bibnamefont{Schoengen}},
  \bibinfo{author}{\bibfnamefont{H.}~\bibnamefont{Doscher}},
  \bibinfo{author}{\bibfnamefont{T.}~\bibnamefont{Hannappel}},
  \bibinfo{author}{\bibfnamefont{B.}~\bibnamefont{Lochel}},
  \bibinfo{author}{\bibfnamefont{M.}~\bibnamefont{Barth}}, \bibnamefont{and}
  \bibinfo{author}{\bibfnamefont{O.}~\bibnamefont{Benson}},
  \bibinfo{journal}{Applied Physics Letters} \textbf{\bibinfo{volume}{97}},
  \bibinfo{pages}{141108} (\bibinfo{year}{2010}), ISSN
  \bibinfo{issn}{00036951}.

\bibitem[{\citenamefont{Nadj-Perge et~al.}(2010)\citenamefont{Nadj-Perge,
  Frolov, Bakkers, and Kouwenhoven}}]{Nadj-Perge:2010}
\bibinfo{author}{\bibfnamefont{S.}~\bibnamefont{Nadj-Perge}},
  \bibinfo{author}{\bibfnamefont{S.~M.} \bibnamefont{Frolov}},
  \bibinfo{author}{\bibfnamefont{E.~P. A.~M.} \bibnamefont{Bakkers}},
  \bibnamefont{and} \bibinfo{author}{\bibfnamefont{L.~P.}
  \bibnamefont{Kouwenhoven}}, \bibinfo{journal}{Nature}
  \textbf{\bibinfo{volume}{468}}, \bibinfo{pages}{1084} (\bibinfo{year}{2010}).

\bibitem[{\citenamefont{Jacques et~al.}(2007)\citenamefont{Jacques, Wu,
  Grosshans, Treussart, Grangier, Aspect, and Roch}}]{Jacques:2007}
\bibinfo{author}{\bibfnamefont{V.}~\bibnamefont{Jacques}},
  \bibinfo{author}{\bibfnamefont{E.}~\bibnamefont{Wu}},
  \bibinfo{author}{\bibfnamefont{F.}~\bibnamefont{Grosshans}},
  \bibinfo{author}{\bibfnamefont{F.}~\bibnamefont{Treussart}},
  \bibinfo{author}{\bibfnamefont{P.}~\bibnamefont{Grangier}},
  \bibinfo{author}{\bibfnamefont{A.}~\bibnamefont{Aspect}}, \bibnamefont{and}
  \bibinfo{author}{\bibfnamefont{J.-F.} \bibnamefont{Roch}},
  \bibinfo{journal}{Science} \textbf{\bibinfo{volume}{315}},
  \bibinfo{pages}{966} (\bibinfo{year}{2007}).

\bibitem[{\citenamefont{Jak\'obczyk}(2002)}]{Jakobczyk:2002}
\bibinfo{author}{\bibfnamefont{L.}~\bibnamefont{Jak\'obczyk}},
  \bibinfo{journal}{Journal of Physics A: Mathematical and General}
  \textbf{\bibinfo{volume}{35}}, \bibinfo{pages}{6383} (\bibinfo{year}{2002}).

\bibitem{Durret:2010} R. Durret, \textit{Probability: Theory and Examples} (Cambridge, 2010).

\bibitem[{\citenamefont{Verstraete et~al.}(2009)\citenamefont{Verstraete, Wolf,
  and Ignacio~Cirac}}]{Verstraete:2009}
\bibinfo{author}{\bibfnamefont{F.}~\bibnamefont{Verstraete}},
  \bibinfo{author}{\bibfnamefont{M.~M.} \bibnamefont{Wolf}}, \bibnamefont{and}
  \bibinfo{author}{\bibfnamefont{J.}~\bibnamefont{Ignacio~Cirac}},
  \bibinfo{journal}{Nat Phys} \textbf{\bibinfo{volume}{5}},
  \bibinfo{pages}{633} (\bibinfo{year}{2009}).

\bibitem[{\citenamefont{Barreiro et~al.}(2011)\citenamefont{Barreiro, Muller,
  Schindler, Nigg, Monz, Chwalla, Hennrich, Roos, Zoller, and
  Blatt}}]{Barreiro:2011}
\bibinfo{author}{\bibfnamefont{J.~T.} \bibnamefont{Barreiro}},
  \bibinfo{author}{\bibfnamefont{M.}~\bibnamefont{Muller}},
  \bibinfo{author}{\bibfnamefont{P.}~\bibnamefont{Schindler}},
  \bibinfo{author}{\bibfnamefont{D.}~\bibnamefont{Nigg}},
  \bibinfo{author}{\bibfnamefont{T.}~\bibnamefont{Monz}},
  \bibinfo{author}{\bibfnamefont{M.}~\bibnamefont{Chwalla}},
  \bibinfo{author}{\bibfnamefont{M.}~\bibnamefont{Hennrich}},
  \bibinfo{author}{\bibfnamefont{C.~F.} \bibnamefont{Roos}},
  \bibinfo{author}{\bibfnamefont{P.}~\bibnamefont{Zoller}}, \bibnamefont{and}
  \bibinfo{author}{\bibfnamefont{R.}~\bibnamefont{Blatt}},
  \bibinfo{journal}{Nature} \textbf{\bibinfo{volume}{470}},
  \bibinfo{pages}{486} (\bibinfo{year}{2011}).

\bibitem[{\citenamefont{Duan et~al.}(2001)\citenamefont{Duan, Lukin, Cirac, and
  Zoller}}]{Duan:2001}
\bibinfo{author}{\bibfnamefont{L.~M.} \bibnamefont{Duan}},
  \bibinfo{author}{\bibfnamefont{M.~D.} \bibnamefont{Lukin}},
  \bibinfo{author}{\bibfnamefont{J.~I.} \bibnamefont{Cirac}}, \bibnamefont{and}
  \bibinfo{author}{\bibfnamefont{P.}~\bibnamefont{Zoller}},
  \bibinfo{journal}{Nature} \textbf{\bibinfo{volume}{414}},
  \bibinfo{pages}{413} (\bibinfo{year}{2001}).

\bibitem[{\citenamefont{Olmschenk et~al.}(2009)\citenamefont{Olmschenk,
  Matsukevich, Maunz, Hayes, Duan, and Monroe}}]{Olmschenk:2009}
\bibinfo{author}{\bibfnamefont{S.}~\bibnamefont{Olmschenk}},
  \bibinfo{author}{\bibfnamefont{D.~N.} \bibnamefont{Matsukevich}},
  \bibinfo{author}{\bibfnamefont{P.}~\bibnamefont{Maunz}},
  \bibinfo{author}{\bibfnamefont{D.}~\bibnamefont{Hayes}},
  \bibinfo{author}{\bibfnamefont{L.-M.} \bibnamefont{Duan}}, \bibnamefont{and}
  \bibinfo{author}{\bibfnamefont{C.}~\bibnamefont{Monroe}},
  \bibinfo{journal}{Science} \textbf{\bibinfo{volume}{323}},
  \bibinfo{pages}{486} (\bibinfo{year}{2009}).

\bibitem[{\citenamefont{Claudon et~al.}(2010)\citenamefont{Claudon, Bleuse,
  Malik, Bazin, Jaffrennou, Gregersen, Sauvan, Lalanne, and
  Gerard}}]{Claudon:2010}
\bibinfo{author}{\bibfnamefont{J.}~\bibnamefont{Claudon}},
  \bibinfo{author}{\bibfnamefont{J.}~\bibnamefont{Bleuse}},
  \bibinfo{author}{\bibfnamefont{N.~S.} \bibnamefont{Malik}},
  \bibinfo{author}{\bibfnamefont{M.}~\bibnamefont{Bazin}},
  \bibinfo{author}{\bibfnamefont{P.}~\bibnamefont{Jaffrennou}},
  \bibinfo{author}{\bibfnamefont{N.}~\bibnamefont{Gregersen}},
  \bibinfo{author}{\bibfnamefont{C.}~\bibnamefont{Sauvan}},
  \bibinfo{author}{\bibfnamefont{P.}~\bibnamefont{Lalanne}}, \bibnamefont{and}
  \bibinfo{author}{\bibfnamefont{J.-M.} \bibnamefont{Gerard}},
  \bibinfo{journal}{Nat Photon} \textbf{\bibinfo{volume}{4}},
  \bibinfo{pages}{174} (\bibinfo{year}{2010}).

\bibitem[{\citenamefont{Schoelkopf and Girvin}(2008)}]{Schoelkopf:2008}
\bibinfo{author}{\bibfnamefont{R.~J.} \bibnamefont{Schoelkopf}}
  \bibnamefont{and} \bibinfo{author}{\bibfnamefont{S.~M.}
  \bibnamefont{Girvin}}, \bibinfo{journal}{Nature}
  \textbf{\bibinfo{volume}{451}}, \bibinfo{pages}{664} (\bibinfo{year}{2008}).

\bibitem[{\citenamefont{Hwang et~al.}(2009)\citenamefont{Hwang, Pototschnig,
  Lettow, Zumofen, Renn, Gotzinger, and Sandoghdar}}]{Hwang:2009}
\bibinfo{author}{\bibfnamefont{J.}~\bibnamefont{Hwang}},
  \bibinfo{author}{\bibfnamefont{M.}~\bibnamefont{Pototschnig}},
  \bibinfo{author}{\bibfnamefont{R.}~\bibnamefont{Lettow}},
  \bibinfo{author}{\bibfnamefont{G.}~\bibnamefont{Zumofen}},
  \bibinfo{author}{\bibfnamefont{A.}~\bibnamefont{Renn}},
  \bibinfo{author}{\bibfnamefont{S.}~\bibnamefont{Gotzinger}},
  \bibnamefont{and}
  \bibinfo{author}{\bibfnamefont{V.}~\bibnamefont{Sandoghdar}},
  \bibinfo{journal}{Nature} \textbf{\bibinfo{volume}{460}}, \bibinfo{pages}{76}
  (\bibinfo{year}{2009}).


\end{thebibliography}

\end{document}